\documentclass[11pt,amsfonts,amsmath]{article}
\usepackage[utf8]{inputenc}
\usepackage{times}
\usepackage{soul}
\usepackage{amssymb}
\usepackage{amsmath}
\usepackage{amsthm}
\usepackage{graphicx}
\usepackage{authblk}
\usepackage{color} 
\usepackage{subfigure}

\graphicspath{{Figures/}}

\newcommand{\R}{{\mathbb{R}}}

\newtheorem{Proposition}{Proposition}

\newcommand{\ModifMK}[1]{\textcolor{black}{#1}}
\newcommand{\ModifFL}[1]{\textcolor{black}{#1}}

\title{Latching dynamics in neural networks with synaptic depression}
\author[1]{Pascal Chossat}
\author[1]{Maciej Krupa}
\author[2]{Fr\'ed\'eric Lavigne}
\affil[1]{\small Universit\'e C\^ote d'Azur Inria CNRS, Mathneuro} 
\affil[2]{\small Universit\'e C\^ote d'Azur CNRS, BCL lab}

\begin{document}
\setstcolor{red}
\maketitle 
\begin{abstract} 
\noindent Priming is the ability of the brain to more quickly activate a target concept in response to a related stimulus (prime). Experiments point to the existence of an overlap between the populations of the neurons coding for different stimuli. Other experiments show that prime-target relations arise in the process of long term memory formation. The classical modelling paradigm is that long term memories correspond to stable steady states of a Hopfield network with Hebbian connectivity. Experiments show that short term synaptic depression plays an important role in the processing of memories. This leads naturally to a computational model of priming, called latching dynamics; a stable state (prime) can become unstable and the system may converge to another transiently stable steady state (target). Hopfield network models of latching dynamics have been studied by means of numerical simulation,
however the conditions for the existence of this dynamics have not been elucidated. In this work we use a combination of analytic and numerical approaches to confirm that latching dynamics can exist in the context of Hebbian learning, however lacks robustness and imposes a number of biologically unrealistic restrictions on the model. In particular our work shows that the symmetry of the Hebbian rule is not an obstruction to the existence of latching dynamics, however fine tuning of the parameters of the model is needed.
\end{abstract}


\section{Introduction}
\ModifFL{Prediction of changes in the environment is a fundamental adaptive property of the brain \cite{Miyashita1988}; \cite{MiyashitaChang1988}; \cite{Miller1999}; \cite{Bunge2003}; \cite{Muhammad2006}). To this aim, the neural mechanisms subtending prediction must activate in memory potential future stimuli on the basis of preceding ones. In nonhuman primates processing sequences of stimuli, neural activity shows two main dynamics triggered by the presentation of the first stimulus (prime) that precede the second stimulus (target). First, some neurons strongly respond to first stimulus and exhibit a retrospective activity at an elevated firing rate after its offset (\cite{Miyashita1988}; \cite{MiyashitaChang1988}; \cite{Fuster1971}). Retrospective activity is considered as a neural mechanism of short-term maintenance of the first stimulus in working memory \cite{Amit1994}; \cite{Goldman-Rakic1995}; \cite{Wang2002}; \cite{Ranganath2005}. Second, some neurons exhibit an elevated firing rate during the delay between the prime and target, i.e. before the onset of the target, and respond strongly to this target \cite{Miyashita1988}; \cite{MiyashitaChang1988}; \cite{Naya2001}; \cite{Naya2003a}; \cite{Naya2003b}; \cite{Yoshida2003}; \cite{Erickson1999}; \cite{Rainer1999}; \cite{Tomita1999}; \cite{Sakai1991}. Prospective activity depends on previous learning of the pairs of prime and target stimuli (\cite{Miyashita1988}; \cite{MiyashitaChang1988}; \cite{Erickson1999}; \cite{Rainer1999}; \cite{Gochin1994}; \cite{Messinger2001}; \cite{Sakai1991}; \cite{Wirth2003}; \cite{Ison2015}) and is considered as a mechanism of prediction of the second stimulus \cite{Wallis2001}; \cite{Wallis2003}; \cite{Brunel2009}. Further, prospective activity of neurons coding for a stimulus is related to response times to process this stimulus when it is presented (\cite{Roitman2002} see \cite{Erickson1999}).
In humans processing sentences, the EEG signal correlates with the level of predictability of words from preceding words \cite{DeLong2005}; \cite{Kutas2010}; \cite{Brothers2015}; \cite{DeLong2014a}; \cite{DeLong2014b} (see \cite{Willems2015} on fMRI and \cite{Ding2015} on MEG signals). The early stages of processing of a word are facilitated when this word is predictable \cite{Lavigne2000}; \cite{McDonald2002}; \cite{DeLong2005}) \ModifMK{leading to} a shorter processing time \cite{Hutchison2014}. This so called priming of a target stimulus by a preceding prime is reliably reported in both human \cite{Meyer1971}; \cite{Neely1991}; \cite{Hutchison2003}; \cite{Lavigne2011}; \cite{Meyer2014} and nonhuman primates \cite{Erickson1999}; see \cite{Brunel2009} for a review. Further, experiments show that the magnitude of priming highly relies on the relation between the two stimuli stored in memory (\cite{Brothers2015}; \cite{VanPetten2014}; \cite{Luka2014}; \cite{Lavigne2012}; \cite{Lavigne2013}). In both human and non-human primates, the relation between two stimuli stored in memory depends on the learned sequences of stimuli (\cite{Spence1990}; \cite{Landauer1998}; \cite{VanPetten2014}).}

Computational modelling \ModifMK{studies} of biologically inspired neural networks \ModifMK{have been carried out in the context of} the dynamics of neural activity in priming protocols used in human and nonhuman primates. Models show that retrospective activity of a stimulus is possible for high values of synaptic efficacy between neurons that are active to code for this stimulus (\cite{Amit1997};\cite{Amit2003}) and that prospective activity of a stimulus not presented is possible for high values of synaptic efficacy between neurons active to code for the first stimulus and neurons active to code for the second stimulus (\cite{Brunel1996}; \cite{Mongillo2003}). On this basis, computational models have shown how a large spectrum of priming phenomena depends on the level of prospective activity of neurons coding for the second stimulus (\cite{Brunel2009}; \cite{Lerner2012}; \cite{Lerner2014}). Taken as a whole, models have emphasized the essential role of the matrix of synaptic efficacies for the generation of specific levels of prospective activity generating specific levels of priming. 

Many neurophysiological studies have described learning at the synaptic level as combinations of long-term potentiation (LTP) and long-term depression (LTD) of synapses (\cite{Hebb1949}; \cite{Bliss1973}; \cite{Bliss1993}; \cite{Kirkwood1994}). On this basis, synaptic efficacy is an essential parameter to code the relation between stimuli in memory (e.g., \cite{Yakovlev1998}; \cite{Weinberger1998}). Further, single cell recordings and local field potentials report that neurons in the macaque cortex respond to several different stimuli (\cite{Rolls1995}; \cite{Tamura2001}; \cite{Tsao2006}) and that a given stimulus is coded by the activity of a population of neurons (\cite{Hung2005}; \cite{Young1992}; \cite{Kreiman2006}). As a result, the information about a specific stimulus is distributed across a pattern of activity of a neural population (\cite{QuianQuiroga2010}; see \cite{QuianQuiroga2016}). Two different patterns of activity corresponding to two stimuli can therefore share some active neurons. Hence, such pattern overlap in the populations responsive to different stimuli can code a relation between these stimuli (\cite{Fujimichi2010}; \cite{QuianQuiroga2012}).

\ModifMK{There have been a number of computational studies focussed on} priming generated by the dynamics of populations of neurons with a distributed coding of the stimuli in attractor network models \cite{Masson1991, Masson1995, Plaut1995, Plaut2000, Mossetal1994}. \ModifFL{When presented with an external stimulus, these attractor networks converge to a stable steady state and do not activate a sequence of patterns. However, \ModifMK{latching dynamics have been described as} the internal activation of a sequence of patterns triggered by an initial stimulus \cite{T05} (see also \cite{Kawamoto1985, Horn1989, Herrmann1993, Kropff2007, Russo2008}). \ModifMK{In a model recently introduced \cite{Lerner2012, Lerner2014}}, several priming effects involved in prediction can be reproduced by latching dynamics that depend on the overlap between the patterns. This is made possible \ModifMK{due} to units that do not maintain constant firing rates, allowing the network to change state instead of converging to a fixed-point attractor. Interestingly, latching dynamics \ModifMK{relies} on the specific neural mechanisms of neural noise and fast synaptic depression. Neural noise is a fundamental property of the brain \cite{Softky1993, Shadlen1998, RDbook}). \ModifMK{One of its functions is} to increase the probability of state transitions in attractor networks \cite{Miller2006, Fiete2014}. However, noise alone does not allow regular sequences of state transitions according to pattern overlap (see \cite{Chaudhuri2016}). Fast synaptic depression reported in cortical synapses \cite{Tsodyks1997} rapidly decreases the efficacy of synapses that transmit the activity of the pre-synaptic neuron. A consequence is that the network cannot sustain a stable regime of activity of the neurons in a given pattern and spontaneously change\ModifMK{s} state. Connectionist models have shown the effects of fast synaptic depression on semantic memory \cite{Huber2003} and on priming \cite{Lerner2012, Lerner2014}. }
\ModifFL{When a stimulus is presented to the network, neurons activated by the stimulus activate each others in a pattern, but fast synaptic depression contributes to their deactivation because they activate \ModifMK{each other} less and less. In the meantime, these neurons \ModifMK{begin} to activate neurons of a different but overlapping pattern, that, because they are less activated, exhibit less synaptic depression at their synapses.} \ModifMK{Before fast synaptic depression takes its effect, the} newly activated neurons can strongly activate their associates in \ModifMK{the new pattern}. \ModifMK{The transition from the old to the new pattern is enabled by the synaptic noise.} \ModifMK{Hence} the combination of neural noise and fast synaptic depression makes  latching dynamics \ModifMK{possible} in attractor neural networks. \ModifFL{However, the precise role of each of these mechanisms in changing the network state are still unclear. Further, the necessary and sufficient pattern overlap for latching dynamics and how it combines with synaptic depression and noise are still unknown.} The aim of the present approach is to analyze the necessary and sufficient conditions of combination of neural noise, fast synaptic depression and overlap for the existence of latching dynamics, using the framework of heteroclinic chains \cite{pcmk}.

The term \textit{heteroclinic chain} refers to a sequence of steady states joined by connecting trajectories. Heteroclinic chains or cycles have been studied in various contexts, including fluid dynamics, population biology, game theory and neuroscience (see \cite{hofbauersigmund, kruparev, rafbiva} for a review), in particular in a model of sequential working memory \cite{rabinovichPRL}. Typically such chains involve states of saddle type, acting as sink for some trajectories and source for other ones. Following the set-up of \cite{Lerner2012, Lerner2014} we use Hopfield networks as attractor network models, however, following \cite{pcmk}, we make a small change in the equation defining the network, in order to ensure that heteroclinic chains can exist in a robust manner. 
Another difficulty is that latching dynamics does not fit into the classical context of heteroclinic chains, as the learned patterns that lose stability due to synaptic depression cannot be seen as states of saddle type.
Hence we need to consider generalized heteroclinic chains given as a sequence of connecting trajectories joining attractors (learned patterns) which become unstable due to a slowly varying variable
(synaptic strength).
The context of heteroclinic chains has the simplicity which allows for the derivation of numerous algebraic conditions that need to be satisfied in order for such chains (and hence latching dynamics) to exist.
Therefore our work leads to a better qualitative and quantitative understanding of latching dynamics, including the role of overlap, synaptic depression, noise and feedback inhibition. 

\section{Materials and Methods}\label{sec-matmet}
\ModifMK{As in} \cite{Lerner2012, Lerner2014}  with somewhat different notations, \ModifMK{the system describing the dynamics of} $N$ neurons is as follows
 \begin{equation}\label{eq-hopfieldnetwork}
 \dot u_i=\frac{1}{\tau}\left(-u_i - \sum_{j=1}^N J_{ij} x_j-I-\lambda\sum_{j=1}^N{x_j} \right),\qquad i=1,\ldots N.
 \end{equation}
where $u_j$ is the \ModifMK{activity variable} membrane potential of neuron $j$ , $I$ is a constant external input, $x_j$ is the firing rate of neuron j, the coefficients $J_{ij}$ express the strength of the excitatory connections from neuron $j$ to neuron $i$
and $\tau$ is the time constant, measured in miliseconds.
 \ModifMK{The terms $-I-\lambda\sum_{j=1}^N{x_j}$ represent inhibition, discussed in more detail below.}
The firing rate is itself a monotonously increasing function of the activity variable with limiting values $0$ and $1$. 
This function is often taken as $x=g(u)=(1+e^{-u/\mu})^{-1}$. \ModifMK{In this work we will use an approximation of $g$, as shown below.}

\ModifMK{System \eqref{eq-hopfieldnetwork} can be expressed in terms of firing rates by means of the transformation $x_i=g(u_i)$. }
\[
\dot x_i=\frac{1}{\tau} x_i(1-x_i)\left(-\mu g^{-1}(x_i)-I-\lambda\sum_{j=1}^N x_j+\sum_{j=1}^N J_{ij} x_j\right),\qquad j=1,\ldots,N.
\]
\ModifMK{Learned patterns are steady state patterns of \eqref{eq-hopfieldnetwork}
of the form $(\xi_1,\ldots,\xi_n)$, $\xi_j=0$ or $1$. 
In order to apply the linearized stability principle we need to be able to evaluate partial derivatives of the right hand side of \eqref{eq-hopfieldnetwork} at the learned patterns. However for such states 
the derivatives do not exist for the particular choice of $g$. 
\ModifFL{This makes it impossible to apply the algebraic method of linearization (computation of eigenvalues of the linearized system) in the current models.} 
We can remedy this using the approach introduced in \cite{pcmk}}
by replacing the function $g^{-1}(x)=\ln x - \ln (1-x)$
by its Taylor expansion $f_q(x)$ at $x=1/2$ up to some arbitrary order $q$. When we let $q$ tend to infinity $f_q$ tends uniformly to $g^{-1}$ in any interval in $(0,1)$. In the following, for simplicity, we take the expansion to first order $f_1(x)=4x-2$, (this corresponds to $q=1$). A different choice of $q$
would not significantly alter our results.  \ModifMK{After renaming the parameters we arrive at the equations}
\begin{equation}\label{eq:firing rate}
\dot x_i=\frac{1}{\tau} x_i(1-x_i)\left(-\mu x_i -I-\lambda\sum_{j=1}^N x_j+\sum_{j=1}^N J_{ij} x_j\right), \qquad i=1,\ldots,N
\end{equation}
The system \eqref{eq:firing rate} has the following fundamental property: any vertex, edge, face or hyperface of the cube $[0,1]^N$ is {\em flow-invariant}: trajectories with starting point in any one of these sets are entirely including in it. This implies in particular that the vertices are equilibria, or steady-states, of \eqref{eq:firing rate}. The vertices have coordinates $0$ (inactive unit) or $1$ (active unit). Hence vertices correspond to patterns for the neural network and whenever a vertex is a stable equilibrium it represents a \ModifMK{learned pattern}.

According to the synaptic depression assumption the coefficients $J_{ij}$ vary in time according to the \st{following} rule:
$$
J_{ij}(t)=J_{ij}^{\rm max} s_j
$$
\ModifMK{where the evolution of the synaptic variable $s_i$ is given as follows \cite{TPM98}}
\begin{equation}\label{eq:syndepr}
\dot s_i=\frac{1-s_i}{\tau_r}-Ux_is_i
\end{equation}
$\tau_r$ and $U$ being \ModifMK{the time constant of the recovery of the synapse and the maximal fraction of used synaptic resources.} 

\ModifMK{Our goal in this work is to investigate latching dynamics in networks which in the absence of  synaptic depression are so-called attractor neural network models
\cite{Masson1995}, \cite{Mossetal1994}, \cite{Plaut1995}. This means that the connectivity matrix $(J^{\rm max})_{i,j=1,\ldots n}$ is derived from a set of learned patterns which must be stable steady states
of the system. Following \cite{Lerner2014} we use the version of the Hebbian rule introduced in \cite{Tsodyks90}. According to this rule the coefficients of the connectivity matrix $(J^{\rm max})_{i,j=1,\ldots n}$
(without synaptic depression) satisfy}
\begin{equation}\label{eq-preheb}
J_{ji}^{\rm max} = \sum_{k=1}^P{\frac{(\xi_i^k-p)(\xi_j^k-p)}{Np(1-p)}},
\end{equation}
\ModifMK{where $\xi^1,\ldots,\xi^P$ are the learned patterns, $N$ is the total number of neurons} and $p$ is the rate of active units in the sparsity of the matrix $J$. 
Let us set $\nu=(Np(1-p))^{-1}$. \ModifMK{We simplify the expression \eqref{eq-preheb} by introducing a change of variables and parameters, see also \cite{Tsodyks90}.}
The rhs of \eqref{eq:firing rate} can be rewritten as 
$$\frac\nu\tau x_i(1-x_i)\left(-\mu/\nu x_i -I/\nu-\lambda/\nu\sum_{j=1}^N x_j+\sum_{j=1}^N J^{\max}_{ij} s_jx_j\right)$$ 
where now
\begin{equation}\label{eq:hebb rule}
J_{ji}^{\rm max} = \sum_{k=1}^P{(\xi_i^k-p)(\xi_j^k-p)}
\end{equation}
Remark that $J^{\max}$ is symmetric, while $J$ at $t>0$ need not be so.
Renaming parameters $\mu/\nu$ as $\mu$ etc, and rescaling time by $t=t'/\nu$ we see that replacing $J_{ij}$ in \eqref{eq:firing rate} by $J_{ji}^{\max}s_j$ with $J_{ji}^{\rm max}$ given by \eqref{eq:hebb rule} does not modify the analysis. In the following we shall assume that $p\ll 1$ (sparse matrix) and replace $p$ by 0 in \eqref{eq:hebb rule}.   Moreover, given that $\nu$ is a constant between $0$ and $1$ and, due to the sparsity of the matrix $J^{\rm max}$, is not particularly close to $0$, we set,
for simplicity, $\nu=\tau$. This choice does not qualitatively alter our results.
Hence the context of our study is system \eqref{eq:firing rate} witth $\tau=1$ and the goal is to find latching dynamics between learned patterns with the connectivity matrix given by \eqref{eq:hebb rule}.
As an intermediate stage of our investigation we will consider systems of the form \eqref{eq:firing rate} with weights that do not satisfy \eqref{eq:hebb rule}.

\ModifMK{We introduce the concept of a {\em heteroclinic chain} and argue that it gives a close approximation of latching dynamics.}
Following \cite{pcmk}, we  \ModifMK{consider connecting trajectories} within edges or faces of $[0,1]^N$ between patterns sharing at least one active unit. Without synaptic depression this is not possible because the patterns in this case are stable equilibria, thanks to the way the symmetric connectivity matrix $J^{\max}$ was built. 
\ModifMK{Hence, for the existence of transitions, we have to assume that at least some of the synaptic variables are less than $1$. We will assume that at time $t=0$ all the synaptic variables have the value $1$
so that all the learned patterns are stable.
Given given a sequence of steady state patterns $\xi^1,\dots,\xi^M$, \ModifMK{$M<P$}, a heteroclinic chain consists of a sequence of connecting trajectories \ModifFL{(dynamic transition patterns)}
between these patterns and of a sequence 
of time instances $0<t_1<\dots < t_M$ such that the \ModifFL{transition} from $\xi_k$ to $\xi_{k+1}$ exists for the coefficients of the connectivity matrix $J_{ij}$ evaluated at $t_k$, that is 
$J_{ij}=J_{ij}^{\rm max} s_j(t_k)$, where $i,j=1,2,\ldots, n$.
If there exists a heteroclinic chain then adding noise will yield latching dynamics. A more detailed discussion of the role of noise is included at the end of this section.} 
The problem can therefore be formulated like this: 
\medskip

\noindent {\bf Problem}: let there be given a sequence of patterns $\xi^1,\dots,\xi^M$, \ModifMK{$M<P$}, where $\xi^k$ and $\xi^{k+1}$ share at least one active unit for all $k=1,\dots, P-1$. Under which conditions does there exist a sequence of \ModifMK{connecting trajectories} $\xi^1\rightarrow\dots\rightarrow\xi^P$, so that a {\em heteroclinic chain} is realized between these patterns?
\medskip

\subsection{\ModifMK{\ModifFL{Effect} of noise} } \ModifMK{ We add noise to \eqref{eq-hopfieldnetwork} in the form of a white noise point process. 
Such a noise term can be thought of as the fluctuation of the firing rate due to the presence of random spikes. 
This means that the noise term has to be suitable adjusted to ensure that negative firing rates or firing rates
greater than $1$ do not arise. In practice, in our simulations we add a noisy perturbation to the initial condition at regular intervals of time, 
making sure that the perturbations are positive for firing rates near $0$
and negative for firing rates near $1$.}

\ModifMK{ The noise is indispensable to transform a heterclinic chain into  `latching dynamics'. When one of the learned patterns becomes unstable due to synaptic depression, 
noise is necessary to make sure that the solution does not linger near the steady state point. If the noise is too large then random spiking gives rise to an increased amount of non-selective inhibition, which may
prevent the existence of latching dynamics.} 

\subsection{\ModifMK{\ModifFL{Effect} of inhibition}}
\ModifMK{The term $-I$ in \eqref{eq:firing rate} corresponds to constant (tonic inhibition). Due to the presence of this term the pattern consisting of all neurons inactive is stable. }

\ModifMK{The term $-\lambda\sum x_i$ is the non-selective inhibition, depending on the activity of the specific neurons. This contribution should be thought of as feedback inhibition: a neuron which
is active excites some interneurons which contribute an inhibitory feedback. We assume that the connectivity matrix is sparse, which is consistent with neurophysiological data 
\cite{Holmgren2003}, as well as with computational models showing that a sparse matrix allows maximal storage capacity \cite{Brunel2016}; \cite{Clopath2012}; \cite{Brunel2004}; \cite{Chapeton2012}; \cite{Clopath2013}. 
Sparse connectivity
and the non-selective inhibition imply that stable patterns contain only a few active neurons. }
\subsection{\ModifMK{Eigenvalue computations}}
The structure of equations \eqref{eq:firing rate} makes the eigenvalues of the system linearized at each steady state pattern lying on a vertex of the hypercube $[0,1]^N$  easy to compute (diagonal Jacobian matrix). Let $\xi=(\xi_1,\dots,\xi_N)$ be a vertex (hence $\xi_j=0$ or $1$), then the eigenvalue at $\xi$ along the coordinate axis $x_k$ has the form 
\begin{equation}\label{eq:eigenvalue}
\sigma_k=(-1)^{\xi_k} \left(-\mu\xi_k-I-\lambda\sum_{j=1}^N \xi_j +\sum_{j=1}^N J^{\max}_{kj}s_j\xi_j\right).
\end{equation}
\ModifMK{The stability condition is now\\ 
\noindent (S)  $\sigma_k<0$ for all $k=1,2,\ldots , n$.\\ 
\noindent Note that this algebraic method would not be available if we had not replaced the transfer function $g$ by its Taylor polynomial.}

\ModifMK{The assumption of sparsity implies that for each $k$ only a few $J^{\max}_{kj}$'s can be non-zero. This means that in a stable pattern only a few $\xi_j$'s can be non-zero, otherwise 
the contribution of the non-selective inhibition would not allow the stability condition to hold.}

\ModifMK{Formula \eqref{eq:eigenvalue} and the stability condition (S) are the tools that will allow us to create conditions for the existence of heteroclinic chains and establish
the role of the overlap between learned patterns. We show that such overlap is needed for the existence of a heteroclinic chain.}

 
\subsection{\ModifMK{Constructing heteroclinic chains}}
\ModifMK{Due to the action of synaptic depression each of the learned patterns in a heteroclinic chain must lose stability due to one or more of the eigenvalues $\sigma_k$ becoming positive.
We assume that no two eigenvalues become positive at the same time, which implies that a noisy trajectory must follow the direction of the unstable eigenvalue.
We will assume that in order to pass from one learned pattern to the next, the trajectory follows the edge corresponding to the unstable eigenvalue to the opposite vertex,
which is a saddle point with a single unstable direction connecting to the next learned pattern in the chain.  
The chain will therefore be a sequence of \textit{elementary chains} consisting of connections with three elements: a learned pattern $\xi^i$ that becomes unstable due to synaptic depression (prime), a \textit{transition}
pattern 
$\hat \xi^{i}$
of saddle type, and the next learned pattern 
$\xi^{i+1}$ (target).
The fact that the transition pattern should be unstable imposes another condition on the eigenvalues. These conditions will be presented in the results section.}

\ModifMK{We argue that an elementary chain is the most likely mechanism of transition from $\xi^i$ to $\xi^{i+1}$. It is certainly the simplest case dynamically.
Any more complicated dynamics would be likely to increase the passage time, so that the target pattern could lose stability due to synaptic depression
before becoming active in the chain. Finally, as will be explained in the sequel, more complicated dynamics would require the existence of additional unstable eigenvalues
leading to additional constraints on the matrix $J^{\rm max}_{ij}$.}
\section{Results}\label{sec-res}
Latching dynamics involves only a few learned patterns, consisting of a small number of active neurons, with significant overlap between the patterns. Hence we expect that there exists a small \textit{subnetwork},
weakly connected to the rest of the network, which supports a heteroclinic chain. The connectivity matrix restricted to this subnetwork is not necessarily obtained from the Hebbian rule.  Based on this argument we break up
the problem into two parts:\\[0.2ex]
- we consider a small network (the prototype of a subnetwork), designing the connectivity matrix so that a heteroclinic chain connecting a priori specified patterns exists,\\[0.2ex]
- we construct a larger network whose connectivity matrix is derived from the Hebbian rule such that the small network is its subnetwork.
  Our construction leads naturally to a matrix with sparse connectivity. It is known that connectivity in the brain is only about 
  $10$ \% \cite{Mason1991}; \cite{Markram1997}; \cite{Sjostrom2001}; \cite{Holmgren2003}; \cite{Thomson2007}; \cite{Lefort2009}, hence our construction is consistent with the biophysical data. \\[0.2ex]
We carry out this procedure for a few examples illustrating the general principle. 

The first step is to derive the algebraic constraints from the eigenvalue conditions  \eqref{eq:eigenvalue} which define the parameter regions where heteroclinic chains could exist.
These conditions are only necessary, in fact our numerics show that heteroclinic chains which follow a prescribed sequence of connections arise in a reliable manner  in yet smaller parameter regions.
 As explained in Section \ref{sec-matmet} a cycle we consider joins a sequence of learned patterns $\xi^1\to\cdots\to\xi^p$ such that each of them has exactly $m$ excited neurons (with entry $1$) and the switching from one pattern to the next corresponds to switching the values in two entries. Possibly after re-arrangement of the indices it is no loss of generality to assume that 
$$
\xi^1=(\overbrace{1,\dots,1}^{m~ times},0,\dots,0)~,~\xi^2=(0,\overbrace{1,\dots,1}^{m~ times},0,\dots,0),\dots,\xi^p=(0,\dots 0,\overbrace{1,\dots,1}^{m~ times},0,\dots,0).
$$
In addition we have $p-1$ transition patterns
$$
\xi^1=(0,\overbrace{1,\dots,1}^{m-1~ times},0,\dots,0),~\xi^2=(0,\; 0, \overbrace{1,\dots,1}^{m-1~ times},0,\dots,0),\dots,\xi^{p-1}=(0,\dots 0,\overbrace{1,\dots,1}^{m-1~ times},0,\dots,0).
$$
We make a simplifying assumption that the entries of $J^{\rm max}$ are $0$ outside of a band around the diagonal of width $2m-1$ (this is consistent with the requirement of the sparsity of the matrix).  We introduce:
\begin{equation}\label{eq-3Lam}
\begin{split}
&\Lambda_{i,k}=\sum_{l=0}^{m-1} J_{k,i+l},\quad 1\le i\le n-m+1,\quad 1\le k\le n\}\\
&\Lambda^{\rm max}=\max_{i,k\not\in \{i,\ldots i+m-1\}} \Lambda_{i,k}\\
& \Lambda^{\rm min}=\min_{i,k\in \{i,\ldots i+m-1\}} \Lambda_{i,k}.
\end{split}
\end{equation}
The requirement that the patterns $\xi^1,\, \ldots\,\,\xi^p$ are stable in the absence of synaptic depression can be expressed, using \eqref{eq:eigenvalue}, by the condition 
\begin{equation}\label{eq-reqgmt}
m\lambda +I>\Lambda^{\rm max}
\end{equation}
\begin{equation}\label{eq-reqlmt}
m\lambda +I<\Lambda^{\rm min}-\mu
\end{equation}
Other types of constraints come from the fact that, in the time interval of  transition from one pattern to the next, the dynamics must approach a transition state from the direction of the prime pattern
and leave in the direction of the target pattern. It means that there is a time instance $\hat t_i$ such that
\begin{equation} \label{eq:condnecessaire1mt}
\sum_{j=i+1}^{m+i-1} J^{\max}_{i,j} s_j(t) < I+(m-1)\lambda < \sum_{j=i+1}^{m+i-1} J^{\max}_{i+m,j} s_j(t),~t\in (\hat t^i_1,\hat t^i_2),
\end{equation}
which implies a weaker condition
\begin{equation} \label{eq:condnecessaire2mt}
I+(m-1)\lambda < \min_{i} \Lambda_{i+m,i+1}.
\end{equation}
The combination of \eqref{eq-reqgmt}, \eqref{eq-reqlmt} and \eqref{eq:condnecessaire2mt} places severe restrictions on the parameters $\lambda$, $I$ and $J^{\rm max}_{ij}$.
Additional constraints can be derived from the fact all the other directions of $\hat \xi_i$ have to be stable, in order to ensure the reliability of the cycle, but we did not explore these conditions here.
For $m=2$ we can use \eqref{eq:condnecessaire1mt} and the fact that there is only one synaptic variable pertaining to $\hat \xi^i$ to obtain the inequality:
$J^{\max}_{i,i+1}<J^{\max}_{i+2,i+1}$. From the symmetry of the connectivity matrix we conclude that $J^{\max}_{i,i+1}<J^{\max}_{i+1,i+2}$. In other words, the elements on the upper diagonal
and the lower diagonal must be increasing. This, combined with  \eqref{eq-reqgmt}, \eqref{eq-reqlmt} and \eqref{eq:condnecessaire2mt}, gives, for $m=2$ 
\begin{equation}\label{eq-condsmt}
\begin{split}
& {\rm (i)}\; J_{i,i+1}<J_{i+1,i+2},\; i=1,\ldots n-1\mbox{ (upper diagonal elements are increasing)}\\
&{\rm (ii)}\; I+\lambda<J_{21}\\
&{\rm (iii)}\; I+2\lambda >J_{m,m+1}\\
& {\rm (iv)}\;  I+2\lambda < \min_i (J_{i,i}+J_{i,i+1}).
\end{split}
\end{equation}
The property of increasing diagonal elements, in practice, prevents the existence of long chains as the large coefficients will activate the corresponding neurons just due to the presence of noise.

We use \eqref{eq-condsmt} to select the parameters for our numerical examples. For the details of the derivation of the algebraic constraints we refer to Appendix \ref{ap-compcoeff}.

Our results show that, given a small network, the parameters have to be tuned quite precisely to obtain a heteroclinic chain. Adding the requirement that the matrix is obtained
using the Hebbian rule gives an even more severe constraint. We have constructed examples of networks supporting heteroclinic chains with each neuron involved in some of the patterns forming
the chain.  In each case the connectivity coefficients we used had larger values than given by the patterns involved in the chain alone. To solve this problem we designed a method
of defining a larger system, with the connectivity matrix of the form
\begin{equation}\label{eq-ext}
\tilde J^{\rm max}=\left (\begin{array}{cc} J^{\rm max} &A\\A^T&B\end{array}\right )
\end{equation}
and added learned patterns which do not participate in the chain but with an overlap with the patterns forming the chain, so that the matrix $\tilde J$ is obtained using the Hebbian rule.
The matrix $A$ consists of many blocks with few non-zero coefficient that are small in comparison to the entries of $J^{\rm max}$.
The matrix $B$ is block diagonal, with the off-diagonal entries in each of the blocks equal to $1$. The added learned patterns must satisfy 
the constraints \eqref{eq-reqgmt} and \eqref{eq-reqlmt} to ensure their stability. 
This way the matrix is sparse (about 25 \% non-zero elements in the example we constructed).
There is no natural algorithm to construct $\tilde J^{\rm max}$, so we refrain from making any further specifications.
We constructed $\tilde J^{\rm max}$ for a specific example, see Appendix \ref{sec-constr}.

%
%
\subsection{A simple example -- an elementary chain}\label{sec-simplex}
\ModifMK{In this section we examine the simplest} case of a network of three neurons \ModifMK{and} two learned patterns
\begin{equation}\label{patterns n=3}
\xi^1=(1,1,0)~\mbox{and}~\xi^2=(0,1,1).
\end{equation}
\ModifMK{This example is the prototype of an elementary chain. }
According to our program stated in Section \ref{sec-matmet} we aim at finding conditions such that a sequence of \ModifMK{connecting} orbits 
\ModifMK{(dynamic transition patterns)} along edges of the cube in $\R^3$ connect these patterns in a chain. This requires the presence of one intermediate equilibrium $\hat\xi^1=(0,1,0)$, which by assumption is not a learned pattern and has an unstable direction along the coordinate which passes from 0 to 1 in the sequence
\begin{equation}\label{chain n=3}
\xi^1\to \hat\xi^1 \to \xi^2.
\end{equation}
Computing the connectivity matrix from the learning rule \eqref{eq:hebb rule} with patterns $\xi^1$ and $\xi^2$ is straightforward and gives
\begin{equation}\label{eq-Jmaxn=3}
J^{\rm max}=\left (\begin{array}{ccc} 1&1&0\\1&2&1\\0&1&1\end{array}\right )
\end{equation}
Applying formula \eqref{eq:eigenvalue} with $N=3$ and $J^{\max}$ given by \eqref{eq-Jmaxn=3} it is easily checked that the two learned patterns have negative eigenvalues, hence are stable, in the absence of synaptic depression, iff $1<I+2\lambda<2-\mu$. This is our first requirement.\\ 
In the remaining of this section $\sigma^i_k$, resp. $\hat\sigma_k$, will denote the eigenvalue at $\xi_i$, resp. $\hat\xi$ along $x_k$.

We now "switch on" synaptic depression. As time elapses, synaptic weights will be modified according to \eqref{eq:syndepr}. For a given pattern (steady-state of \eqref{eq:firing rate}) on a vertex of the cube $[0,1]^3$) the evolution of the synaptic variable $s_i$ ($i=1, 2$ or 3) can be of two types: (i) if $x_i=0$ the value $s_i=1$ is a steady-state of \eqref{eq:syndepr}; (ii) if $x_i=1$ then $s_i$ decreases monotonically towards the limit value $S=(1+\tau U)^{-1}$ which is the steady-state of \eqref{eq:syndepr} at $x_i=1$. 

Let us describe in terms of the eigenvalues associated with each pattern, the scenario which would lead to the expected heteroclinic chain. \\
First, the weakening of the synaptic variable $s_1$ should modify the stability of the learned pattern $\xi_1$ in such a way that a trajectory will appear connecting $\xi_1$ to the intermediate $\hat\xi$ along the first coordinate $x_1$. This entails that the eigenvalue $\sigma^1_1$ at $\xi_1$ (eigenvalue along coordinate $x_1$), which initially is negative, becomes positive after some time. This is a kind of {\em dynamic bifurcation}, in which time plays the role of a bifurcation parameter through the evolution of the synaptic variables. Simultaneously we want the eigenvalue $\hat\sigma_1$ at $\hat\xi$ become negative at finite time so that this state is attracting along direction $x_1$. By \eqref{eq:eigenvalue} a sufficient condition for this is $-I-\lambda+J^{\max}_{12}S<0$. \\
The second step is to check conditions which would allow for the existence of a trajectory connecting $\hat\xi$ to $\xi_2$ along the edge with coordinate $x_3$. This requires that the eigenvalue $\hat\sigma_3$ be positive after some time (possibly already at $t=0$), hence by \eqref{eq:eigenvalue}, $-I-\lambda+J^{\max}_{32}S>0$. The two conditions we just derived are incompatible with matrix \eqref{eq-Jmaxn=3}, so we can conclude that this matrix does not admit the heteroclinic chain \eqref{chain n=3}.

\begin{figure}[htbp]
\begin{center}
\includegraphics[width=7cm]{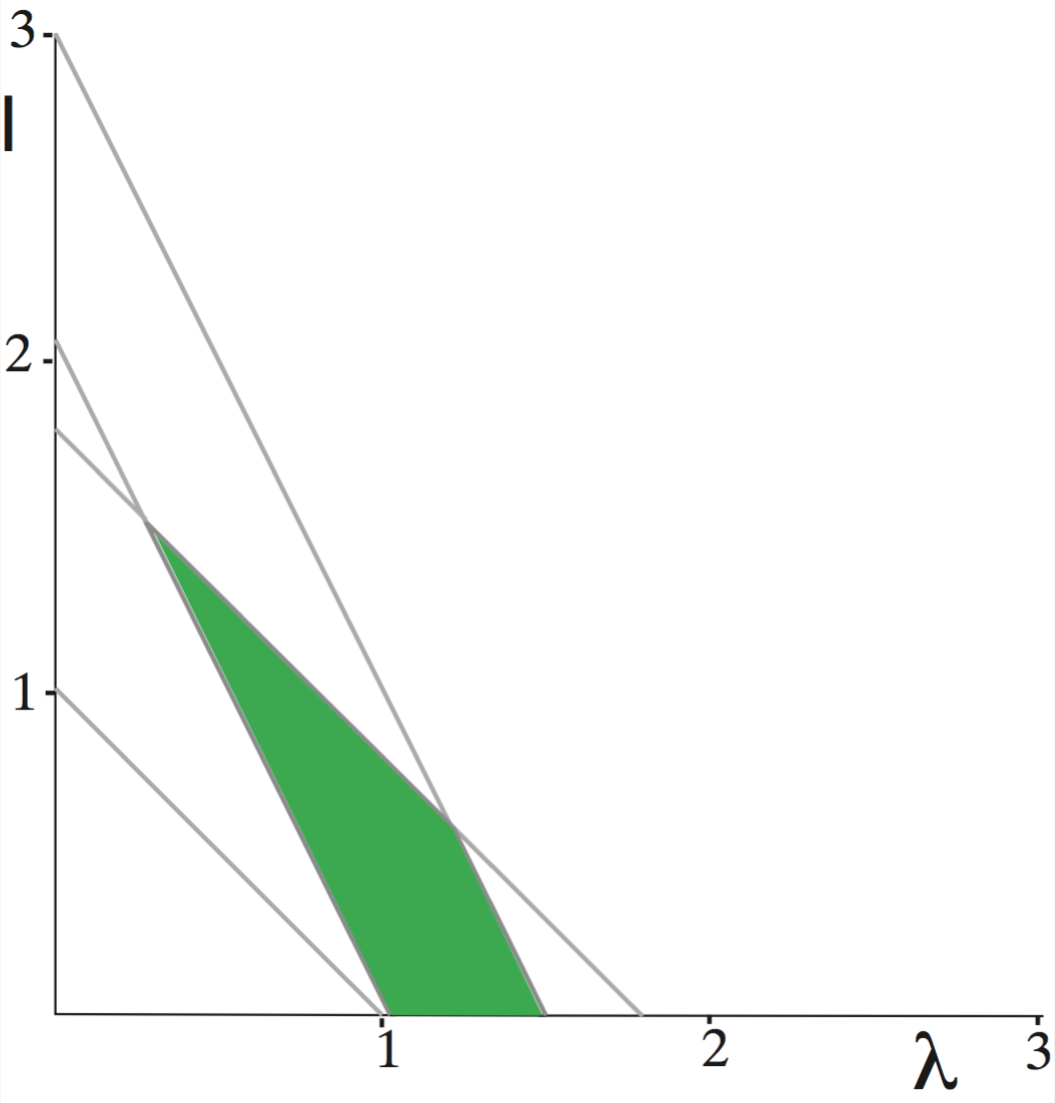}
\caption{{\small The domain of values $(\lambda,I)$ which allow for existence of a heteroclinic dynamics with matrix \eqref{eq-Jmaxn=3 version 2} and $S=1/1.4$.}}
\label{fig:conditions graphiques}
\end{center}
\end{figure}

From the above discussion we can infer that a necessary condition for the existence of a chain \eqref{chain n=3} is that $J^{\max}_{32}>J^{\max}_{12}$. Since $J^{\max}$ is symmetric this means that its upper diagonal must have strictly increasing coefficients.
In addition to these conditions we also request that: (i) $\xi_1$ be "more" stable in the $x_2$ direction than in the $x_1$ direction, i.e. $\sigma^1_2<\sigma^1_1<0$ at $t=0$, so that a trajectory starting close to this equilibrium will first destabilze in the $x_1$ direction (it may of course not destabilize at all); (ii) the $x_2$ direction at $\hat\xi$ is stable; (iii) $\xi_2$ is stable when $t$ large enough. This imposes additional conditions on the coefficients of $J^{\max}$. Let us show that the matrix
\begin{equation}\label{eq-Jmaxn=3 version 2}
J^{\rm max}=\left (\begin{array}{ccc} 2&1&0\\1&3&2\\0&2&2\end{array}\right )
\end{equation}
satisfies all these conditions and hence admits the heteroclinic chain \eqref{chain n=3}. Of course this is an ad'hoc construction, however we shall show in Appendix 
\ref{sec-constr} that \eqref{eq-Jmaxn=3 version 2} is a submatrix of the connectivity matrix of a subnetwork of a large sparse network under Hebbian rule (see \eqref{eq-ext}). \\
For \eqref{eq-Jmaxn=3 version 2} the eigenvalues computed using \eqref{eq:eigenvalue} are as follows:
\begin{equation}
\begin{array}{cccccccc}
\sigma^1_1 &=& I+2\lambda-(2s_1+s_2-\mu)  &  \sigma^1_2 &=& I+2\lambda-(s_1+3s_2-\mu) & \sigma^1_3 &= -I-2\lambda+2s_2 \\
\hat\sigma_1 &=& -I-\lambda+s_2  &  \hat\sigma_2 &=& I+\lambda-(3-\mu) & \hat\sigma_3 &= -I-\lambda+2s_2 \\
\sigma^2_1&=& -I-2\lambda+s_2 &  \sigma^2_2 &=& I+2\lambda-(3s_2+2s_3-\mu)  & \sigma^2_3 &= I+2\lambda-(2s_2+2s_3-\mu)
\end{array}
\end{equation}
From this we obtain the following conditions:
\begin{enumerate}
\item $2<I+2\lambda<3$ (stability of the learned patterns in absence of synaptic depression),
\item $3S<I+2\lambda$ ($\sigma^1_1$ becomes $>0$ in finite time),
\item $1<I+\lambda$ ($\hat\sigma_1<0$),
\item $I+\lambda<2S$ ($\hat\sigma_3$ becomes $>0$ in finite time),
\item $I+2\lambda<2S+2$ ($\sigma^2_3<0$).
\end{enumerate}
These conditions are all satisfied if $\tau$ and $U$ are chosen such that $2/3<S$ and the point $(\lambda,I)$ lies in the triangle bounded by the lines $I>0$, $2<I+2\lambda$ and $I+\lambda<2S$. The value of $S=(1+\tau U)^{-1}$ being given, these conditions can be conveniently represented graphically. Figure \ref{fig:conditions graphiques} shows an example with $\tau=100$ and $U=0.004$.  

\begin{figure}[htbp]
\begin{center}
\includegraphics[width=11cm]{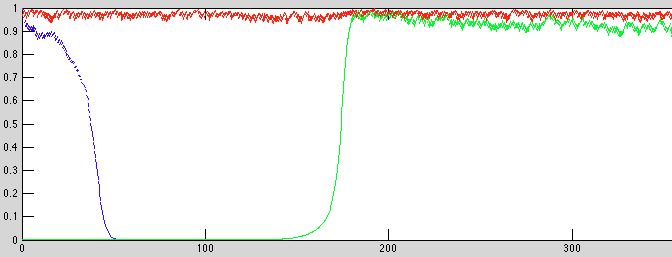}
\includegraphics[width=11cm]{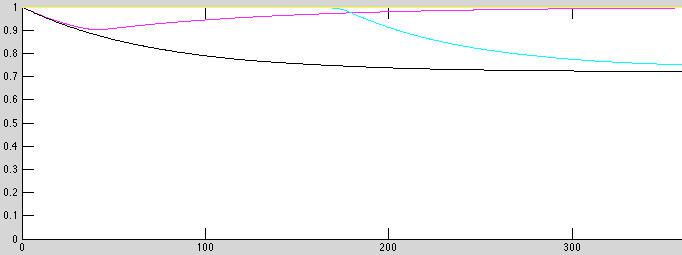}
\caption{{\small Top: time series of $x_1$ (blue), $x_2$ (red) and $x_3$ (green). Bottom: time series of $s_1$ (magenta), $s_2$ (black) and $s_3$ (cyan).}}
\label{fig:timeseries N=3}
\end{center}
\end{figure}

Let us illustrate numerically this result. We have integrated the equations \eqref{eq:firing rate} and \eqref{eq:syndepr} with $N=3$ and coefficient values $\tau=100$, $U=0.004$, $I=0.15$ and $\lambda=1.2$. Figure \ref{fig:timeseries N=3} shows a time series of $x_I(t)$ (upper figure) and $s_i(t)$ (lower figure) with an initial condition starting close to $\xi_1$. 
In order to observe the transitions \eqref{chain n=3} in reasonable time we have incorporated a noise in the code, in the form of a small random deviation from initial condition on the $x_i$ variables at each new integration time (simulation with Matlab).
\section{Numerical examples}\label{sec-num}
\subsection{A case with five neurons and the extended network of 61 neurons}\label{sec-5261}
\begin{figure}[htbp]
\begin{center}
\includegraphics[scale=0.6]{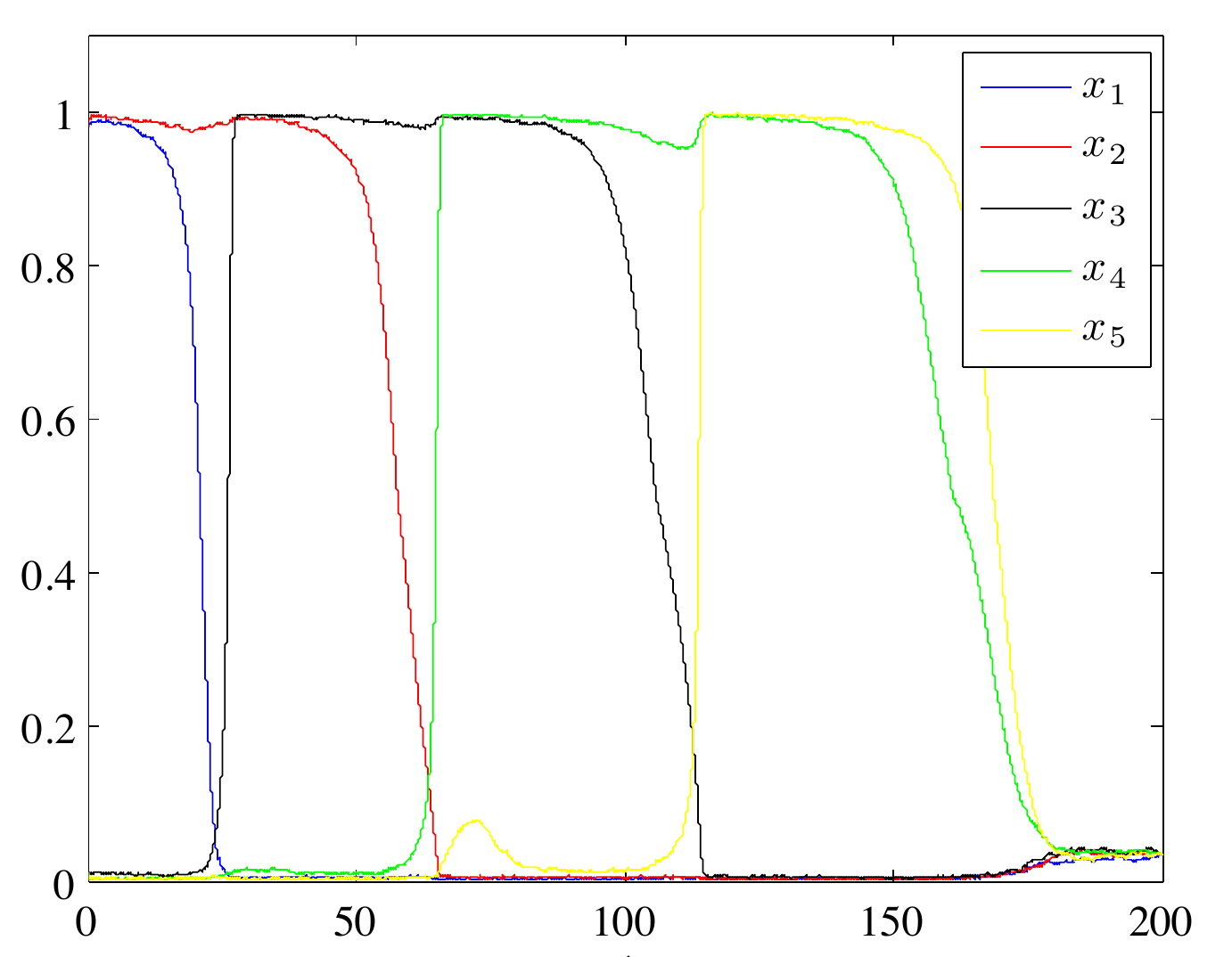}
\caption{{\small The chain of four patterns in the network of 5 neurons  with the transition matrix \eqref{eq-Jmax5p} and the parameters
$I=0.3$, $\lambda=3.4$, $\mu=3.1$, $\tau_r=400$ and $U=0.01$.}}
\label{fig-chain5261}
\end{center}
\end{figure}

We consider
\begin{equation}\label{eq-Jmax5p}
J^{\rm max}=\left (\begin{array}{ccccc} 9&3&0&0&0\\3&10&5&0&0\\0&5&11&6&0\\0&0&6&11&7\\0&0&0&7&11\end{array}\right ),
\end{equation}
with $I=0.3$, $\lambda=3.4$, $\mu=3.1$, $\tau_r=400$ and $U=0.01$. This matrix and these parameter values meet all conditions for the existence of a heteroclinic chain joining the patterns $\xi^1=(1,1,0,0,0),\dots,\xi^4=(0,0,0,1,1)$, however \eqref{eq-Jmax5p} does not follow from Hebbian rule. In figure \ref{fig-chain5261} we show a simulation of a chain of four states existing for the above parameters.

We extended this network to a network of 61 neurons with the connectivity matrix satisfying the Hebbian rule using the approach described in Section \ref{sec-res}. The details of the construction
can be found in Appendix \ref{sec-constr}. Figure \ref{fig-chain61} shows a simulation for the required chain \eqref{eq-1st4}.
\begin{figure}[htbp]
\begin{center}
\includegraphics[scale=0.6]{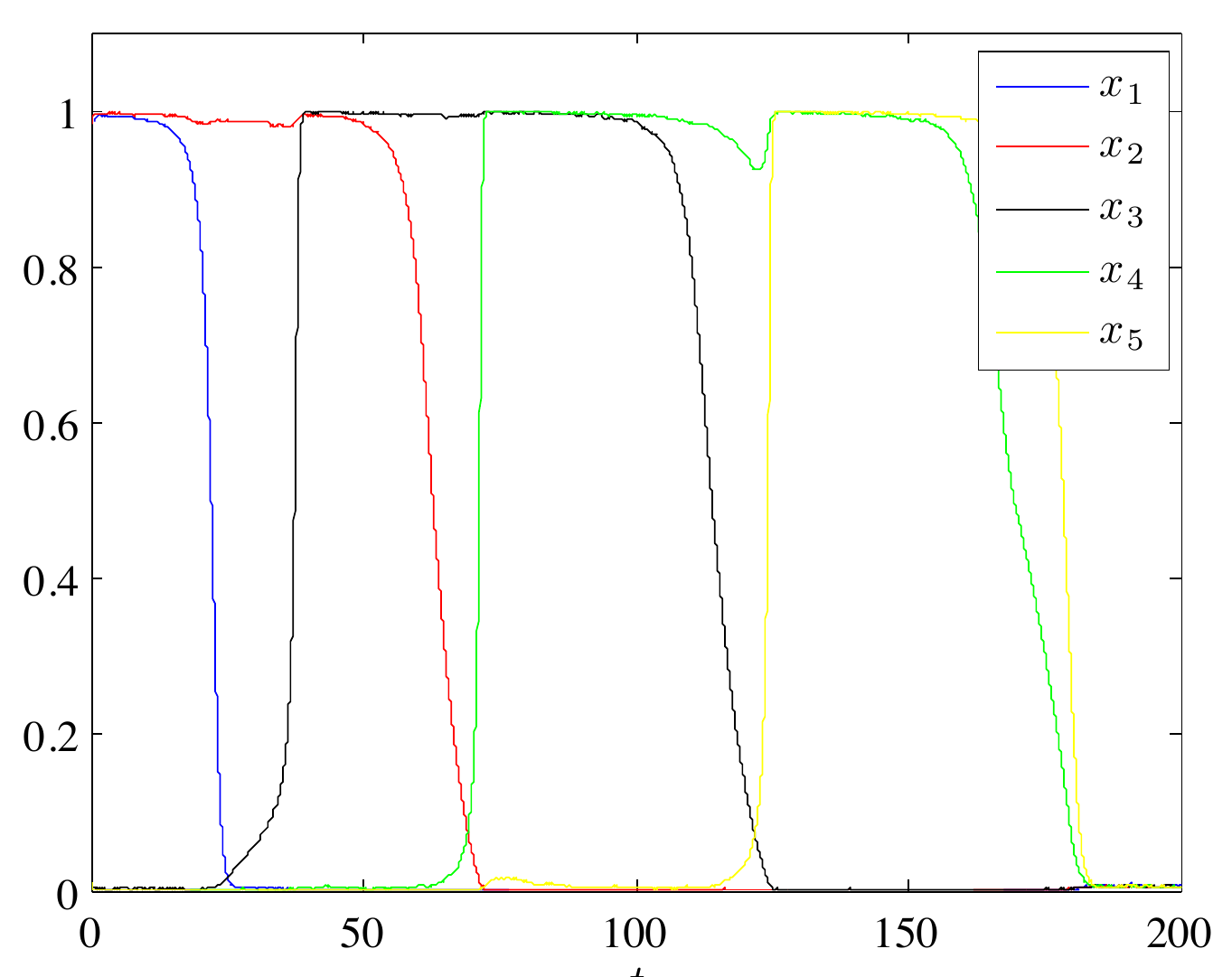}
\caption{{\small The chain of four patterns in the network of 61 neurons obtained by extending the five neuron example with the transition matrix \eqref{eq-Jmax5p}.
The parameter settings are as the same as in the simulation of Figure \ref{fig-chain5261}.}}
\label{fig-chain61}
\end{center}
\end{figure}
\subsubsection{Sparsity of the extended network}
 Electrophysiological studies suggest that connectivity in the brain is sparse with only approximately 10\% of pairs of neurons connected \cite{Mason1991}; \cite{Markram1997}; \cite{Sjostrom2001}; \cite{Holmgren2003}; \cite{Thomson2007}; \cite{Lefort2009}. In the extended network obtained in this section and in Appendix \ref{sec-constr} the 'emptiness' of the matrix (fraction of zero-weight synapses) is above 75\%, which is consistent with neurophysiological data as well as with computational models showing that a sparse matrix allows maximal storage capacity \cite{Brunel2016}; \cite{Clopath2012}; \cite{Brunel2004}; \cite{Chapeton2012}; \cite{Clopath2013}. The present results show that sparsity is necessary not only to improve storage capacity (ensure the stability of learned patterns)  but also to enable the sequential activation of patterns. Indeed, in the case of Hebbian learning considered here, heteroclinic chains involving patterns defined by the activity of neurons e.g. 1-6 are possible only if the synaptic matrix obeys conditions on the efficacies along the subdiagonal. These conditions depend in turn on the role of additional neurons (n) among a large number of 'non-coding neurons' taken into account in the learning equation. This is possible under conditions of sparse coding of the patterns.
\subsubsection{Reliability of the chain}\label{sec-relchain}
\begin{figure}[htbp]
\begin{center}
\includegraphics[scale=0.55]{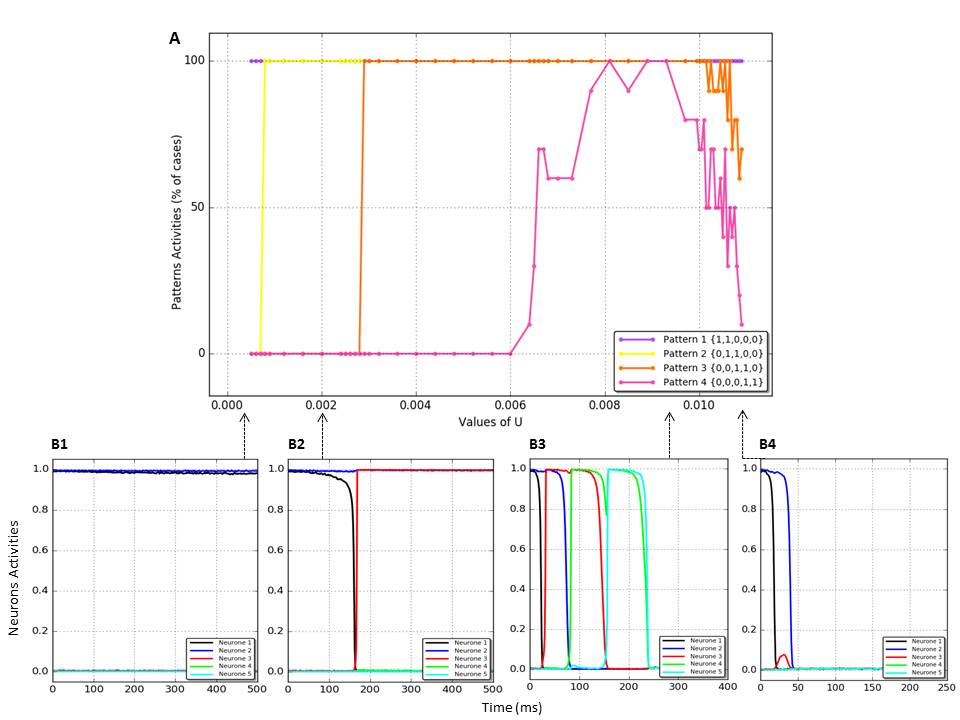}
\caption{{\small Panel A. Reliability of the chain of four patterns in the network of 61 neurons as a function of U. Pannels B1-4. Activities of neurons coding for the patterns in the chain as a function of time for four representative values of U.}}
\label{fig-CarlosFred}
\end{center}
\end{figure}

\ModifFL{We have computed the reliability of the chain with the prescribed sequence $(1,1,0,0,0)\to (0,1,1,0,0)\to (0,0,1,1,0)\to (0,0,0,1,1)$ in the extended network of 61 neurons, with respect to the parameter $U$ (maximal fraction of used synaptic resources). For parameter values distributed in the interval $0.0005\le U\le 0.110$ 
we performed 10 simulations with the same initial conditions for each of the chosen parameter values. Figure \ref{fig-CarlosFred} shows the distribution, for different values of U, of the proportion of the simulations for which a given pattern was activated in the chain. }

\ModifFL{
The top panel A shows that the activation of the full length chain (all of the four successive patterns) is possible for a limited range of values of U (from .0081 to .0093; pannel B3). Even within this range of values of U and with fixed parameter values, the chain does not always fully develop on every simulation due to the presence of noise. Further, for values of U lower or higher than this range, the chain does not fully develop for two different reasons. On the one hand, when U decreases, the length of the activated chain decreases because the network stays in the state corresponding to the first pattern (pannel B1) or to the second pattern with slow transition time (pannel B2). On the other hand, when U increases, the length of the activated chain decreases because the network does not stay in the state corresponding to the first but does not activate the second pattern either, ending in a state where no neuron is activated (pannel B4).
Taken as a whole, these results show that the value of U determines the reliability of the chain in terms of number of patterns activated, with the patterns occurring later in the chain less likely to be activated. Further, the value of U also determines the state of the network after the first pattern (ranging from this first pattern or the different following patterns (low values of U) to an absence of activity of the neurons (high values of U). For the optimal range of values of U, the reliability of the chain is maximal but not perfect due to the presence of noise.}

\subsection{A chain connecting six patterns with $m=2$}
In this section we present an example of a longer chain involving six neurons and 5 patterns.
We do not construct the extension to a large network with a Hebbian matrix. This would be possible using the approach outlined in Section \ref{sec-res} and 
carried out in detail for the example of five neurons in Appendix \ref{sec-constr}.  However the matrix we consider has larger entries than the one of the preceding section (five neurons), which
implies that a larger extended network would be needed.

\begin{figure}[htbp]
\begin{center}
\includegraphics[width=15cm]{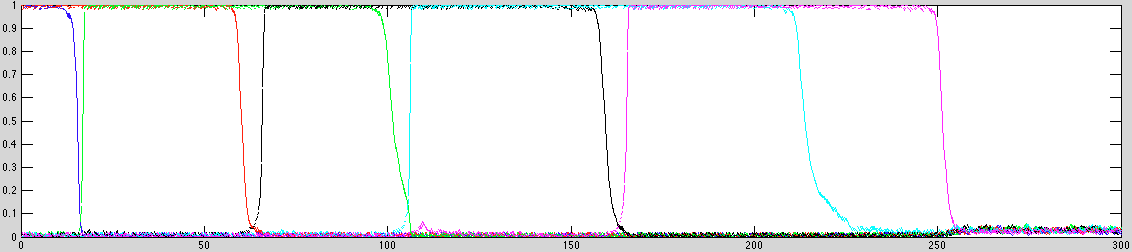}
\includegraphics[width=15cm]{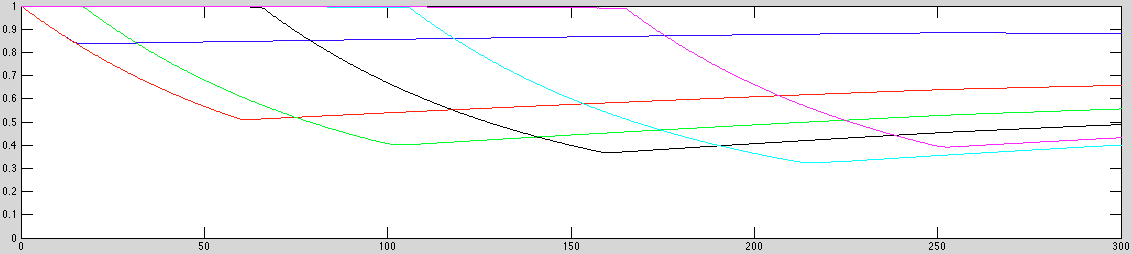}
\caption{{\small Chain of five patterns with 6 neurons and $m=2$. Top panel showing $x_j$ and the bottom panel $s_j$. Color code: blue=$x_1$, red=$x_2$, green=$x_3$, black=$x_4$, cyan=$x_5$, magenta=$x_6$. Same code for $s_j$. }}
\label{fig:6xs}
\end{center}
\end{figure}

We consider the following connectivity matrix:
\begin{equation}\label{eq-Jmax6}
J^{\rm max}=\left (\begin{array}{cccccc} 13&6&0&0&0&0\\6&14&13&0&0&0\\0&13&16&14&0&0\\0&0&14&20&15&0\\0&0&0&15&20&16\\0&0&0&0&16&20\end{array}\right )
\end{equation}
This matrix satisfies all the necessary conditions as explained in Section \ref{sec:condnecessaire} with $m=2$, based on the learned patterns $\xi^1=(1,1,0,0,0,0)$,  $\xi^2=(0,1,1,0,0,0)$,
$\xi^3=(0,0,1,1,0,0)$, $\xi^4=(0,0,0,1,1,0)$ and $\xi^5=(0,0,0,0,1,1)$.
For the choice of parameters  $I=0.48$,  $\lambda=8$,  $\mu=1.2$,  $\tau_r=600$ and $U=0.012$ and for a range of noise amplitudes this matrix gives the following heteroclinic chain/latching dynamics: \\
Starting with initial condition close to $\xi^1$, the dynamics visits successively $\xi^2,\dots,\xi^5$, the transition form $\xi^i$ to $\xi^{i+1}$ passing through the intermediate (not learned) state $\hat\xi^i$ with only one excited neuron at rank $i+1$, see Fig. \ref{fig:6xs}.
Observe that as long as a variable $x_j$ is "large" (close to 1) the corresponding synaptic variable $s_j$ decreases until $x_j$ comes close to 0. Then $s_j$ increases, according to the time evolution driven by \eqref{eq:syndepr}. 
\subsection{A case of a shared neuron with m=3}
\begin{figure}[htbp]
\begin{center}
\includegraphics[width=15cm]{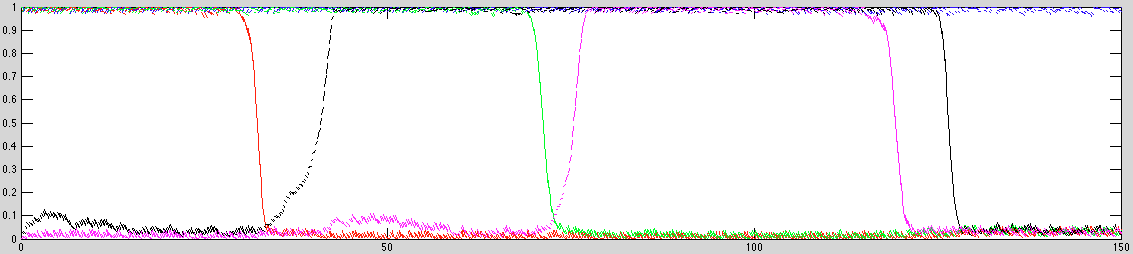}
\includegraphics[width=15cm]{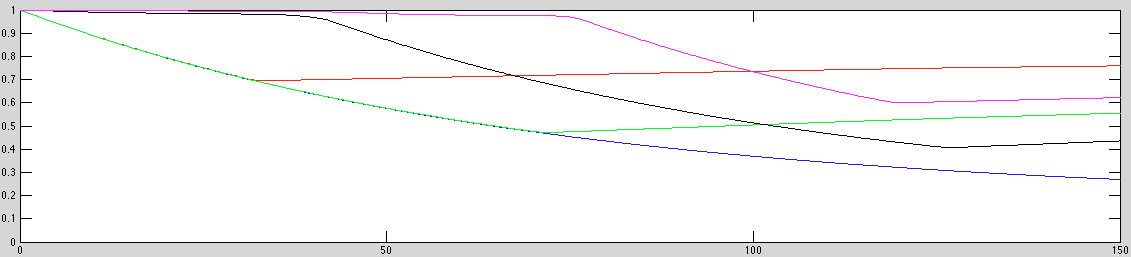}
\caption{{\small Chain of three patterns with 5 neurons, m=3 and one shared neuron. Color code: blue=$x_1$, red=$x_2$, green=$x_3$, black=$x_4$, magenta=$x_5$. Same code for $s_j$. }}
\label{fig:shared}
\end{center}
\end{figure}

\ModifMK{This example gives a different option for the neural coding of items.}
Thus far the principle of our model has been that each item (e.g. the prime or target) is coded by a pattern of activity of all neurons in the network. As a consequence only one item can be 'activated' at a given time in a heteroclinic chain. \ModifFL{The simplest case is when two patterns are activated in succession: the pattern coding for the prime followed by a pattern coding for the target \cite{Masson1991, Masson1995, Plaut1995, Lerner2012}. In this case  either the prime or the target is 'activated' at a given time. Such priming mechanism can account for neuronal activities recorded in nonhuman primates in priming protocols where a prime is related to a single target. In that case priming relies on the successive activation of the presented prime and of the predicted target \cite{Erickson1999, Mongillo2003}. However, in human studies priming is reported not only for targets directly related to the prime (Step 1 targets), but also for targets indirectly related to the prime through a sequence of one (Step 2 targets) or two (Step 3 targets) intermediate associates of the prime that are activated after the prime and before the target (e.g. \cite{Brunel2009} for a review). Such indirect priming has been accounted for by network models in which Step 1, Step 2 and Step 3 associates to the primes were coded by neural populations that can be activated simultaneously \cite{Brunel2009}. The present model of heteroclinic chains is of particular interest to account for the sequential activation of items involved in step priming. However, in priming studies the prime can still be reported by participants after processing of the target \cite{Dark1988}. This suggests that the prime must be available in working memory at the end of the activation of the sequence of Step associates, that is neurons coding for the prime must be active at the end of the heteroclinic chain. The possibility to activated the prime in after several associates have been activated in a chain is no reproduced by models of priming based on latching dynamics \cite{Lerner2012, Lerner2014}. In the present model, a way to for neurons coding for the prime to be actvated at the end of the heteroclinic chain is simply to consider that a pattern (attractor state) does not correspond to a single item, but rather corresponds to several items each corresponding to the activity of a subgroup of neurons. In that case the activity of a neuron would correspond to the average activity of a population of neurons coding for and item \cite{Ison2015}. Such population coding is consistent with recent models of priming in the cerebral cortex \cite{Brunel2009, Lavigne2011, Lavigne2014}. Intuitively, the first pattern in the chain codes for the prime only while the next pattern 2 in the chain codes for the combination of the prime and of the Step 1 target together, the pattern 3 codes for the Prime, Step1 target and Step 2 target, 
and so on}. 
\ModifMK{This way the population coding for the prime would be active throughout the entire computation.}
In this section we present an example of a system of five neurons with three active neurons in each pattern and one neuron present in each pattern.
For this we use the following connectivity matrix:
\begin{equation}\label{eq-Jmax5}
J^{\rm max}=\left (\begin{array}{ccccc} 12&2&4&4&4\\2&6&3&0&0\\4&3&6&4&0\\4&0&4&7&6\\4&0&0&6&7\end{array}\right )
\end{equation}
$\tau_r=400$, $x_{\rm max}=1$, $U=0.012$. For this system, for the choice of the parameters $I=0.5$, $\lambda=2.8$ and $\mu=1$
we find by simulation a chain joining $\xi_1=(1,1,1,0,0)$, $\xi_2=(1,0,1,1,0)$, $\xi_3=(1,0,0,1,1)$, $\hat\xi_1=(0,1,1,0,0)$ and $\hat\xi_2=(0,0,1,1,0)$.
The time series of the solution is shown in Figure \ref{fig:shared}.

 The above analysis of the network behavior shows that heteroclinic chains can develop in the case where a neuron (i.e. a population coding for the prime) is active for all the successive patterns in the chain. In other words the overlap between populations coding for different items is such that a subgroup of neurons coding for an item (e.g. the prime) can remain activated while another subgroup can be deactivated as the chain progresses. The network is able to keep previous stimuli activated (e.g. a prime) while at the same time it can activate a sequence of items (i.e. associates to the prime) that can be predicted on the basis of the prime. 
The compatibility between changing patterns in the chain and stable activity of a neuron or population of neurons allows to account for two fundamental properties exhibited by the brain, within a unified model. Due to the structure of the overlap in the coding of the memory items, this example combines the population coding used classically in models of priming in the cerebral cortex, in which a given item is coded by a given population of neurons, and the distributed coding used in Hopfield types models of priming, in which a given item is coded by the pattern of activity of all neurons in the network. In this way the present model aims at unifying our understanding of the coding of items in memory and of priming processes between these items.



\section{Discussion}

The present study provides the first analysis of the sufficient conditions for heteroclinic chains of overlapping patterns in the case of Hebbian learning. 
Heteroclinic chains \ModifMK{closely approximate \textit{latching dynamics}, hence
they} are good candidates to account for priming processes reported in human and nonhuman primates. Here priming-based prediction is seen as the activation - by a pattern presented to the network - of a pattern not (yet) presented. \ModifFL{Within this framework, heteroclinic chains account for the activation or inhibition of neurons so that the network codes for the `target' pattern before its actual presentation, under conditions of overlap with the pattern coding for the `prime' pattern. Heteroclinic chains account for different dynamics of activity of neurons reported in nonhuman primates during the delay between the prime and target: some neurons active for the prime and for the target remain active (pair coding neurons \cite{Miyashita1988};\cite{MiyashitaChang1988}), some neurons active for the prime but not for the target are deactivated and some neurons not active for the prime but active for the target exhibit an increased activity during the delay, which corresponds to prospective activity \cite{Naya2001, Naya2003a, Naya2003b, Yoshida2003, Erickson1999}}. 
The model of latching dynamics has been \ModifMK{adapted to allow for the existence of} heteroclinic chains, by replacing the equation for the membrane potential by the equation for the firing rate, with the nonlinearity replaced by its polynomial approximation (to arbitrary order), so that the dynamics is well defined even when the firing rate takes its minimal (0) or maximal (1) values \cite{pcmk}. \ModifMK{In the modified model we were able to} identify \ModifMK{some of the restrictions} 
\ModifMK{imposed on the network by the requirement of} the existence of heteroclinic chains. From a \ModifMK{modelling} perspective, this is a step in bridging the gap between Hopfield-type models of priming and cortical network models of priming. The present model combines several properties that could serve future applications of the model to a better understanding of the relation between perturbation of priming processes reported in schizophrenia (e.g. \cite{Lavigne2008, Lerner2012b, Rolls2008}). The model exhibits latching dynamics reported to account for priming processes and their perturbations, and it calculates spike rates of neurons coding for items in terms of overlap between related populations. \ModifFL{The present model provides a mathematically tractable description of the reliability of sequences of patterns used to model priming in non-human primates and in human. It makes it possible to better understand pathologies of priming, such as Alzheimer disease or schizophrenia, by analyzing the reliability of sequences as a function of network parameters usually considered as subtending perturbations of priming (noise, dopaminergic activity, synaptic connectivity); \cite{Lerner2012b, Lavigne2008, Rolls2008, Brunel2009}.}

\subsection{Predictions on neural activities in priming}
\ModifFL{The present model makes predictions regarding the possibility for the prime to remain activated (remembered) or not (forgotten) as the chains progesses. In the network, activating patterns coding for several Step 1, Step 2, etc targets would make difficult the simultaneous persistent activity of the prime, due to retroactive interference based on inhibition generated by the `step' targets (\cite{Lavigne2011}. The corresponding experimental prediction that could be tested in priming experiments is that the activity of neurons coding for the prime would decrease when successive targets are predicted in memory even though they are not actually presented. This could be visible in nonhuman primates on a decrease of the retrospective activity of neurons coding for the prime when a series of `Step' targets is predicted. The behavioral counterpart in humans would be a decrease in the reportability of the prime when the length of the sequence of targets to predict increases.}

\subsection{Asymmetry of priming and of synaptic efficacies}
Brunel \cite{Brunel2016} recently pointed \ModifMK{out} the possibility that the optimal synaptic matrix depends on the constraint imposed on the network, either storing patterns as stable states or storing patterns to be activated in sequences. The present results show that heteroclinic chains are possible with symmetric matrices built through Hebbian learning and \ModifMK{specify} the necessary conditions for sequences to arise in the network. Although asymmetric connections can improve the ability of the network to activate sequences of patterns, they are not a necessary condition. However, the present results also show that the conditions for heteroclinic chains impose strong constraints on the structure of the synaptic matrix, suggesting that although symmetric weights can be optimal for storage capacity, they are not an optimal solution for the activation of sequences. If the network codes for a given pattern 1 at a given time, the activation of the next pattern 2 in a chain requires that two given neurons $i$ and $j$ activate each other or not depending on their state within each pattern. For example, $i$ but not $j$ can be activated in 1, and the opposite in 2. Hence for an optimal sequence $1\rightarrow 2$, $i$ should activate $j$ but $j$ should not activate $i$.  
The present results show that the symmetry of the weights can be compensated by the sparsity of the network at the expense of an increasing number of neurons \ModifFL{necessary to code for the patterns. Even though asymmetric heteroclinic chains are possible with symmetric synaptic efficacies, further analysis of heteroclinic chains with asymmetric learning rules would bring new evidence on the specific role of asymmetric weights on the required level of sparsity and on the reliability of latching dynamics.}

\appendix
\section{Derivation of the algebraic constraints} \label{ap-compcoeff}
\subsection{Stability of learned patterns in the absence of synaptic depression}\label{sec:condnecessaire}
The example with three neurons ($n=3$) in Sec. \ref{sec-simplex} corresponds to the ideal situation where there is an open domain in the space of parameters of the problem, such that the heteroclinic chain connecting learned patterns along edges of the hypercube $[0,1]^n$ indeed exists. Ideally we would like to derive a set of necessary and sufficient conditions in a general setting, however this seems to be much more difficult to obtain with larger networks and longer chains. Nevertheless dynamics following edges of the hypercube that  \ModifMK{visit learned patterns in an a priori specified sequence, with the coefficients 
derived using the conditions presented in Section \ref{sec-res}, can be observed as shown in Section \ref{sec-num}}. In this section we show that conditions on $J^{\max}$ and coefficients in the equations can be expressed, which strongly restrict the choice of these quantities.

\begin{figure}[htbp]
\begin{center}
\includegraphics[width=10cm]{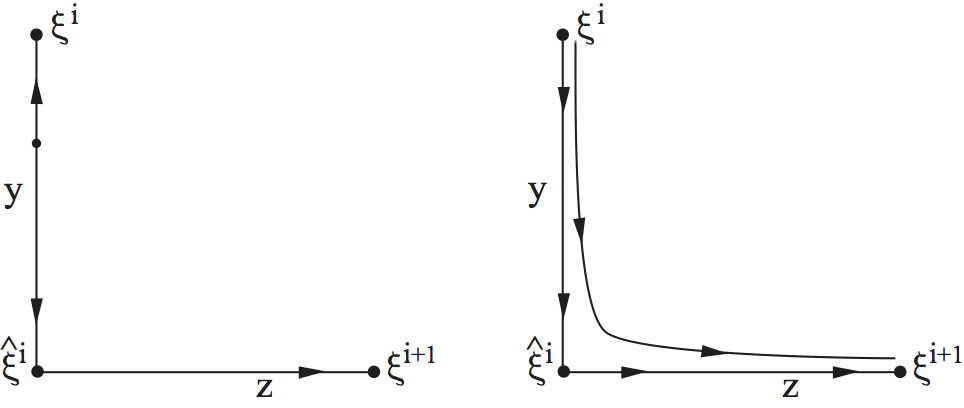}
\caption{{\small Connections from $\xi^i$ to $\xi^{i+1}$ through $\hat\xi^i$}. Left: $t<t_i$, right: $t>t_i$.}
\label{fig:edgeconnexions}
\end{center}
\end{figure}

Recall that in our setting a learned pattern is a stable (when no synaptic depression is present) vertex equilibrium $\xi=(\xi_1,\dots,\xi_n)$ where each $\xi_j=0$ or $1$. Let us consider a sequence of learned patterns $\xi^1\to\cdots\to\xi^p$ such that each of them has exactly $m$ excited neurons (with entry $1$) and the switching from one pattern to the next corresponds to switching values in two entries. Possibly after re-arrangement of the indices it is no loss of generality to assume that 
$$
\xi^1=(\overbrace{1,\dots,1}^{m~ times},0,\dots,0)~,~\xi^2=(0,\overbrace{1,\dots,1}^{m~ times},0,\dots,0),\dots,\xi^p=(0,\dots 0,\overbrace{1,\dots,1}^{m~ times},0,\dots,0).
$$
For any pattern $\xi^i$ ($i<p$) in the above sequence, let $y$ be the coordinate corresponding to its first non-zero entry and z be the coordinate corresponding to the first 0 entry after the sequence of 1's. For example in $\xi^1$, $y=x_1$ and $z=x_{m+1}$. 
We also note $\hat\xi^i$ the vertex equilibrium which makes the connection from $\xi^i$ to $\xi^{i+1}$: its coordinates are those of $\xi^i$ except the $i$-th entry which is 0 instead of 1. By assumption these intermediate states $\hat\xi^i$ are not learned patterns and are saddles for the dynamics of \eqref{eq:firing rate} with all $s_i$ fixed to 1. We specifically require the eigenvalue at $\hat\xi^i$ along $y$ be negative and the eigenvalue along $z$ be positive. \\
Initially the learned patterns $\xi^i$ are stable equilibria of eqs \eqref{eq:firing rate} and synaptic variables $s_i$ are assigned their maximal value $1$. As time evolves the values $s_j(t)$ corresponding to excited neurons ($x_j(t)>0$) decrease and the eigenvalues \eqref{eq:eigenvalue} are modified.  
We expect that after some time $t_i$ the eigendirection $y$ at $\xi^i$ becomes unstable and a heteroclinic connection is established along this edge to $\hat\xi^i$. The synaptic variables play the role of dynamically evolving parameters. If they did not depend on time, the situation could be described as a classical bifurcation: as long as the eigenvalues at $\xi^i$ and $\hat\xi^i$ are both negative an unstable equilibrium exists on the edge joining the two states. When the synaptic variables associated with the excited neurons decrease this equilibrium moves towards $\xi^i$ and when it merges in it, the heteroclinic connection is established (see figure \ref{fig:edgeconnexions}). 
We also expect $\hat\xi^i$ to have an unstable eigendirection along $z$ so that a heteroclinic connection exists from $\hat\xi^i$ to $\xi^{i+1}$. These scenarios can be explicitely described by writing the equations restricted to the edges with variable coordinates $y$ and $z$.  \\
This process should repeat itself from $i=1$ to $i=p-1$.

 We make a simplifying assumption that the entries of $J^{\rm max}$ are $0$ outside of a band around the diagonal of width $2m-1$.  
 It follows from this assumption and from \eqref{eq:eigenvalue} that when $s_j=1$ the eigenvalues at each pattern $\xi^i$ have the following form
\begin{eqnarray}\label{eq:eig xi^i} 
\sigma^i_k &=& -I-m\lambda +\sum_{j=i}^{i+m-1} J^{\max}_{k,j} ~\mbox{when}~k<i ~\mbox{or}~ k>i+m-1 \\
\sigma^i_k &=& \mu+I+m\lambda -\sum_{j=i}^{i+m-1} J^{\max}_{k,j} ~\mbox{when}~ i\leq k\leq i+m-1
\end{eqnarray}
We require all these eigenvalues to be negative. 
 This gives two conditions, which we now specify. Let
\[
\Lambda_{i,k}=\sum_{l=0}^{m-1} J_{k,i+l},\quad 1\le i\le n-m+1,\quad 1\le k\le n\}.
\]
We define 
\[
\Lambda^{\rm max}=\max_{i,k\not\in \{i,\ldots i+m-1\}} \Lambda_{i,k},\quad \Lambda^{\rm min}=\min_{i,k\in \{i,\ldots i+m-1\}} \Lambda_{i,k}.
\]
The two requirements become
\begin{equation}\label{eq-reqg}
m\lambda +I>\Lambda^M
\end{equation}
\begin{equation}\label{eq-reql}
m\lambda +I<\Lambda^m-\mu
\end{equation}
Note that $\Lambda_{i,k}\neq 0$ if $|k-i|<m$ and equals $0$ otherwise. Note also that the computation of $\Lambda^m$ involves non-diagonal elements only.  Hence by making the diagonal elements large enough, it is possible to ensure that
conditions \eqref{eq-reqg} and \eqref{eq-reql} hold. 

\subsection{Necessary conditions on $J^{\max}$ for the transition from one pattern to the next}

We analyze the transition at $\hat\xi^i$ between $\xi^i$ and $\xi ^{i+1}$.
Let $\sigma^i_y, \hat\sigma^i_z$ be the eigenvalues along eigendirections $y$ and $z$ at $\hat\xi^i$ and time $t$. From \eqref{eq:eigenvalue} we have
\begin{eqnarray} \label{eq:eig hat xi^i}
\hat\sigma^i_y &=& -I-(m-1)\lambda +\sum_{j=i+1}^{m+i-1} J^{\max}_{i,j} s_j(t) \\
\hat\sigma^i_z &=& -I-(m-1)\lambda +\sum_{j=i+1}^{m+i-1} J^{\max}_{i+m,j} s_j(t)
\end{eqnarray}
We want that during an interval of time $0<\hat t^i_1<t<\hat t^i_2$ $\hat\sigma^i_y$ becomes negative while $\hat\sigma^i_z$ is positive, which give the condition
\begin{equation} \label{eq:condnecessaire1}
\sum_{j=i+1}^{m+i-1} J^{\max}_{i,j} s_j(t) < I+(m-1)\lambda < \sum_{j=i+1}^{m+i-1} J^{\max}_{i+m,j} s_j(t),~t\in (\hat t^i_1,\hat t^i_2)
\end{equation}
The quantity $S=(1+\tau U)^{-1}$ is the minimal value that can be attained by the synaptic variables $s_j$. Recall the assumption that the elements of $J^{\rm max}$ are $0$ outside a strip of width $2m-1$
about the diagonal. Then   \eqref{eq:condnecessaire1} implies a weaker but simpler condition:
\begin{equation} \label{eq:condnecessaire2}
S \max_i\; \Lambda_{i,i+1} < I+(m-1)\lambda < \min_{i} \Lambda_{i+m,i+1}.
\end{equation}
When $m=2$   \eqref{eq:condnecessaire1} implies the following simple condition:
\begin{Proposition}
If $m=2$, then \eqref{eq:condnecessaire1} implies that $J^{\max}_{i,i+1}<J^{\max}_{i+2,i+1}=J^{\max}_{i+1,i+2}$ (by symmetry of the matrix). Therefore the upper diagonal of $J^{\max}$ has coefficients with increasing values (and the subdiagonal too).
\end{Proposition}
\noindent Indeed, this condition does not hold for matrix \eqref{eq-Jmaxn=3} while matrix \eqref{eq-Jmaxn=3 version 2} satisfies it. \\
For the case $m=2$, in addition to the requirement that the off diagonal coefficients must increase, we deduce the following simple conditions:\\
\begin{equation}\label{eq-conds}
\begin{split}
&{\rm (i)}\; I+\lambda<J_{21}\\
& {\rm (ii)}\; I+\lambda> S J_{m,m+1}\\
&{\rm (iii)}\; I+2\lambda >J_{m,m+1}\\
& {\rm (iv)}\;  I+2\lambda < \min_i (J_{i,i}+J_{i,i+1}).
\end{split}
\end{equation}
A geometric representation of \eqref{eq-conds} similar to Fig. \ref{fig:conditions graphiques} could be constructed.
It is straightforward to verify that the conditions are satisfied for the coefficients of \eqref{eq-Jmax6}. Note also that the conditions derived are quite restrictive, so that the coefficients of $J^{\rm max}$ have to be very carefully chosen (learnt).

Our interpretation of the principle of Hebbian learning is that learned patterns should be dynamically stable for the network (in the absence of synaptic depression). In this case latching dynamics, or heteroclinic chains cannot exist,
as the dynamics would be attracted to the stable state, representing a simple concept, and no passage to the next concept would be possible. The role of synaptic depression is
to destabilize the stable learned patterns and thus enable the passage from one context to the next.

Note that the Hebbian matrix is symmetric. The symmetry of the synaptic weights is compatible with the optimal storing of patterns in sparse networks \cite{Brunel2016}. However, the present results show that a symmetric matrix of efficacies must obey precise conditions that are rather restrictive not sufficient for the existence of heteroclinic chains. In this work we present some of these strong constraints and show that combined effects of and noise and synaptic depression, lead to the existence of heteroclinic chains. It is tempting to think that non-symmetric learning rules would lead to a more robust and predictable presence of heteroclinic chains. This is a topic for future research.

\section{Implementation of the conditions on a sparse network with Hebbian rule}\label{sec-constr}
A straightforward calculation shows that none of the matrices $J^{\rm max}$ considered in the prevoius sections are obtained using the Hebbian rule from the patterns
$\xi^1,\dots,\xi^p$. It is natural to ask if these matrices can be derived using formula  \eqref{eq:hebb rule} by adding more neurons and more learned patterns, so that the patterns $\xi^1,\dots,\xi^p$ 
forming the heteroclinic chain are a part of a larger ensemble of learned patterns and the extended connectivity matrix has the form 
Our fundamental assumption (postulate) throughout this work is that a learned pattern must be stable in the absence of synaptic depression. 
In this section we show that the answer is positive in two cases we have investigated: the case with three neurons and $J^{\rm max}$ as in Section \ref{sec-simplex}, and a more involved case with five neurons and four patterns in the heteroclinic chain. We suspect these results can be extended to more general situations but this is yet to be proved.
\subsection{The case with $n=3$}
Note that the condition $J^{\rm max}_{32}>J^{\rm max}_{21}$ derived in Section \ref{sec-simplex} would imply 
that there must exist a learned pattern different than $\xi^2$, in which neurons 2 and 3 are active and neuron 1 is not. This is obviously not possible
with three neurons. In this section we show that it is possible to obtain a connectivity matrix in a network of six neurons that is derived using
the Hebbian rule and has the matrix \eqref{eq-Jmaxn=3 version 2} in the top left corner. We write
\[
\xi^1=(1,1,0,0,0,0) \mbox{ and } \xi^2=(0,1,1,0,0,0).
\]
These are the patterns $\xi^1$ and $\xi^2$ generalized to the network of six neurons. In addition we consider the patterns:
\begin{equation}\label{eq-extendedpatterns}
\xi^{3}=(0,1,1,1,0,0),\quad  \xi^{4}=(1,0,0,0,0,1),\quad \xi^{5}=(0,0,0,1,1,0),\quad \xi^{6}=(0,0,0,0,1,1).
\end{equation}
Note that $\xi^3$ satisfies precisely the condition stated above, i.e. neurons 2 and 3 are active and neuron 1 is not.
The purpose of adding pattern  $\xi^4$ is to make $ J^{\rm max}_{11}=2$. The role of patterns $\xi_5$ and $\xi_6$ is to ensure the
stability of the added patterns as steady states of \eqref{eq:firing rate} without synaptic depression.

One can easily verify that the matrix 
\begin{equation}\label{eq-Jmaxn=3extended}
J^{\rm max}=\left (\begin{array}{cccccc} 2&1&0&0&0&1\\1&3&2&1&0&0\\0&2&2&1&0&0\\0&1&1&2&1&0\\0&0&0&1&2&1\\1&0&0&0&1&2\end{array}\right )
\end{equation}
is obtained by using formula \eqref{eq:hebb rule} from the patterns $\xi^1$, $\xi^2$, $\xi^{3}$, $\xi^{4}$, $\xi^{5}$ and $\xi^{6}$.
In addition we verify that the learned patterns $\xi^1$, $\xi^2$, $\xi^{3}$, $\xi^{4}$, $\xi^{5}$ and $\xi^{6}$
are stable steady states of \eqref{eq:firing rate} without synaptic depression, ie where $J_{ij}$ have been replaced by $J_{ij}^{\rm max}$.
We perform the computation for the pattern $\xi^{3}$ and show that the additional condition $3\lambda+ I<4$ is needed. Clearly there exist choices of $\lambda$ and $I$ so that this condition as well as 
conditions 1.-5. of Section \ref{sec-simplex} all hold.

\noindent Proceeding analogously as in Section \ref{sec-simplex} we obtain the following eigenvalues:
\begin{eqnarray*}
\sigma^{3}_1&=&-I-3\lambda+s_2,\; \sigma^{3}_2=I+3\lambda-(3s_2+2s_3+s_4-\mu),\; \sigma^{3}_3=I+3\lambda-(2s_2+2s_3+s_4-\mu),\\
\sigma^{3}_4&=&I+3\lambda-(s_2+s_3+2s_4-\mu),\;\sigma^{3}_5=-I-3\lambda+s_4,\; \sigma^{3}_6=-I-3\lambda
\end{eqnarray*}
These expressions are all negative for $s_2=s_3=s_4=1$ and $\mu=0$ if $1<3\lambda+I<4$. The calculations for the other patterns are similar as the calculations in Section \ref{sec-simplex}
and lead to the condition $2<2\lambda+I<3$, i.e. condition 1. from Section \ref{sec-simplex}. 

\subsection{A case with five neurons}
Recall the example of five neurons supporting a chain of four elements that was introduced in Section \ref{sec-5261}
\begin{equation}\label{eq-Jmax5pA}
J^{\rm max}=\left (\begin{array}{ccccc} 9&3&0&0&0\\3&10&5&0&0\\0&5&11&6&0\\0&0&6&11&7\\0&0&0&7&11\end{array}\right ),
\end{equation}
with $I=0.3$, $\lambda=3.4$, $\mu=3.1$, $\tau_r=400$ and $U=0.01$. This matrix and these parameter values meet all conditions for the existence of a heteroclinic chain joining the patterns $\xi^1=(1,1,0,0,0),\dots,\xi^4=(0,0,0,1,1)$, however \eqref{eq-Jmax5pA} does not follow from Hebbian rule \eqref{eq:hebb rule}. In figure \ref{fig-chain5261} we show a simulation of a chain of four states existing for the above parameters.

Our goal is to extend it to a Hebbian matrix using a similar approach as shown in the example with $n=3$. The existence of such an extension was announced in Section \ref{sec-5261},
here we carry out the construction in detail.

We consider a network with $61$ neurons, with patterns of activity represented by vectors with $61$ components. which we introduce below. 
We first introduce some notation: let $e_j\in\R^{61}$ be the vector with $1$ in the $jth$ spot and $0$'s elsewhere. Similarly, $e_{j, k}$ is the vector with $1$'s in the $j$th and $k$th spots and $0$'s elsewhere,
 $e_{j, k, l}$ the vector with $1$'s in the $j$th, $k$th and $l$th spots and $0$'s elsewhere, etc.

The extended network has the four learned patterns to be joined by a chain, which extend the patterns $\xi^1,\dots,\xi^4$:
\begin{equation}\label{eq-1st4}
e_{1,2},\quad e_{2,3},\quad e_{3,4}\quad\mbox{and}\quad e_{4,5}.
\end{equation}
In addition it has the patterns of the form
\begin{equation}\label{eq-extragroup}
\begin{split}
&e_{1,2, j},\mbox{ $j=6,7$ } ,\quad e_{2,3, j},\mbox{ $j=20, 21, 22, 23$ }\\
&e_{3,4,j},\mbox{ $j=34,35,36,37,38$ },\quad  e_{4,5,j},\mbox{ $j=48, 49, 50, 51, 52, 53$ }\\
&e_{1,j},\mbox{ $j=6,7$ } ,\quad e_{1,7,j},\mbox{ $j=20, 21, 22, 23$ }\\
&e_{2,j},\mbox{ $j=6, 7, 8, 9$ },\quad  e_{3,j},\mbox{ $j=20, 21, 22, 23$ }\\
&e_{4,j},\mbox{ $j=34,35,36$ },\quad  e_{5,j},\mbox{ $j=48,49,\ldots, 57$ .}
\end{split}
\end{equation}
and the patterns of the form
\begin{equation}\label{eq-stabs}
\begin{split}
& e_{i,j},\; i=6,7,\ldots , 19,\; j=6,7,\ldots , 19,\\
& e_{i,j},\; i=20,21,\ldots , 33,\; j=20,21,\ldots , 33,\\
& e_{i,j},\; i=34,35,\ldots , 47,\; j=34,35,\ldots , 47,\\
& e_{i,j},\; i=48,49,\ldots , 61,\; j=48,49,\ldots , 61.
\end{split}
\end{equation}
The role of patterns \eqref{eq-extragroup}  is to strengthen the weights in $J^{\rm max}$. The patterns \eqref{eq-stabs} play the role of
stabilizing the patterns \eqref{eq-extragroup}.

The Hebbian matrix derived from all the patterns patterns is:
\begin{equation}
\tilde J^{\rm max}=\left (\begin{array}{ccccc} J^{\rm max}&A_1&A_2&A_3&A_4\\
A_1^T&L_1&M&0_{14\times 14}&0_{14\times 14}\\
A_2^T&M^T&L_2&0_{14\times 14}&0_{14\times 14}\\
A_3^T&0_{14\times 14}&0_{14\times 14}&L_3&0_{14\times 14}\\
A_4^T&0_{14\times 14}&0_{14\times 14}&0_{14\times 14}&L_4
\end{array}\right ),
\end{equation}
where $A_1$, $A_2$, $A_3$ and $A_4$ are $14\times 5$ matrices given by
\begin{equation}\label{eq-A1s}
\begin{split}
&A_1=\left (\begin{array}{cccccccc}2&2&4&0&0&0&\ldots &0\\
                                                      2&2&1&1&1&0&\ldots &0\\
                                                       0&&\ldots&\ldots&&&&0\\
                                                       0&&\ldots&\ldots&&&&0\\
                                                       0&&\ldots&\ldots&&&&0\\
                                                                                                     \end{array} \right )
 A_2=\left (\begin{array}{cccccccc}1&1&1&1&0&\ldots&&0\\
                                                      1&1&1&1&0&&\ldots &0\\
                                                       2&2&2&2&0&&\ldots&0\\
                                                       0&&\ldots&\ldots&&&&0\\
                                                       0&&\ldots&\ldots&&&&0\\
                                                                                                     \end{array} \right )\\
&A_3=\left (\begin{array}{cccccccc}0&&&\ldots&\ldots& &&0\\
                                                       0&&\ldots&\ldots&&&&0\\
                                                      1&1&1&1&1&0&\ldots &0\\
                                                       2&2&2&1&1&0&\ldots&0\\
                                                       0&&\ldots&\ldots&&&&0\\
                                                                                                     \end{array} \right )   
A_4=\left (\begin{array}{cccccccccccc}0&&&&& &&&&&&\ldots 0\\
                                                       0&&&&&&&&&&&\ldots 0\\
                                                       0&&&&&&&&&&&\ldots 0\\
                                                      1&1&1&1&1&1&0&&&&&\ldots 0\\
                                                       2&2&2&2&2&2&1&1&1&1&0&\ldots0\\
                                                                                                     \end{array} \right )                                                                                                                                                                                                  
\end{split}
\end{equation}
\begin{equation}\label{eq-M}
M=\left (\begin{array}{ccccccc}&&& 0_{2\times 14} &&&\\
                                                 1&1&1&1&0&\dots&0\\
                                                &&& 0_{11 \times 14} &&& 
                                                \end{array} \right )  
\end{equation}
and $L_1$, $L_2$, $L_3$ and $L_4$ are defined as follows. Let $N$ be the $14\times 14$ matrix with $0$s on the diagonal and $1$'s off the diagonal.
Then
\begin{equation}
\begin{split}
& L_1={\rm diag}\{15,\, 15,\,  17,\, 13,\, 13,\, 13,\,\ldots, 13\}+N\\
& L_2={\rm diag}\{17,\, 17,\,  17,\, 17,\, 13,\, 13,\,\ldots, 13\}+N\\
& L_3 ={\rm diag}\{15,\, 15,\,  15,\, 14,\, 14,\, 13,\,  13,\,\ldots, 13\}+N\\
& L_4={\rm diag}\{15,\, 15,\,  15,\, 15,\, 15,\, 15,\,14,\, 14,\, 14,\, 14,\, 13,\, 13,\, 13,\, 13\}+N.
\end{split}
\end{equation}
To see that the added patterns are stable, note that the non-diagonal elements outside of $J^{\rm max}$ are all less or equal to $2$, with the exception of $J^{\rm max}_{1,7}=4$.
Note that $2\lambda + I>4$ and $3\lambda +I>6$. Hence the directions not corresponding to one of the active elements must be stable.
For the patterns with two active elements the weakest eigenvalue occurs for the patterns  $e_{1,j}$, $j=6,7$ and equals $-11+2\lambda+I+\mu=-0.8<0$. All the other
eigenvalues are more negative. For the patterns with three active neurons the situation is similar, the patterns $e_{1,2, j}$, $j=6, 7$, give the weakest eigenvalue equal to
$-(9+3+2)+(3\lambda +I +\mu )=-0.4$. Figure \ref{fig-chain61} shows a simulation for the required chain \eqref{eq-1st4}. 

\section*{Acknowledgements}
This work was partially funded by the ERC advanced grant NerVi number 227747.

\end{document}